\documentclass[twocolumn,fleqn,runningheads]{svjour3}
\usepackage{graphicx,amsmath,psfrag,amssymb,makeidx}
\usepackage{booktabs,rotating,subfigure}
\usepackage{url} 
\usepackage{pstricks,pst-node} 
\usepackage{natbib}

\renewcommand{\theta}{\vartheta}
\newcommand{\revision}[1]{ #1}

\begin{document}
\title{Prediction of lethal and synthetically lethal knock-outs in regulatory
networks}
\author{Gunnar Boldhaus, Florian Greil, Konstantin Klemm}
\authorrunning{G.\ Boldhaus, F.\ Greil, K.\ Klemm}

\institute{
G.\ Boldhaus \at Bioinformatics Group, Institute for Computer Science,
Universit\"at Leipzig, H\"artelstra\ss{}e 16-18, 04107 Leipzig, Germany
 \and
F. Greil \at Alfred Wegener Institute for Polar and Marine Research,
Am Handelshafen 12, 27570 Bremerhaven, Germany
 \and
K.\ Klemm \at Bioinformatics Group, Institute for Computer Science,
Universit\"at Leipzig, H\"artelstra\ss{}e 16-18, 04107 Leipzig, Germany
 \email{klemm@bioinf.uni-leipzig.de}
}

\maketitle

\begin{abstract}
The complex interactions involved in regulation of a cell's
function are captured by its interaction graph. More often
than not, detailed knowledge about enhancing or suppressive
regulatory influences and cooperative effects is lacking
and merely the presence or absence of directed interactions
is known.  Here we investigate to which extent such reduced
information allows to forecast the effect of a knock-out or
a combination of knock-outs. Specifically we ask in how far
the lethality of eliminating nodes may be predicted by
their network centrality, such as degree and betweenness,
without knowing the function of the system. The function is
taken as the ability to reproduce a fixed point under a
discrete Boolean dynamics. We investigate two types of
stochastically generated networks: fully random networks
and structures grown with a mechanism of node duplication
and subsequent divergence of interactions. On all networks
we find that the out-degree is a good predictor of the
lethality of a single node knock-out. For knock-outs of
node pairs, the fraction of successors shared between the
two knocked-out nodes (out-overlap) is a good predictor of
synthetic lethality. Out-degree and out-overlap are locally
defined and computationally simple centrality measures that
provide a predictive power close to the optimal predictor.
\end{abstract}


\section{Introduction} \label{sec:Intro} 

The survival, functioning and growth of a living cell is based on a large set of
interdependent biochemical interactions. Interaction networks \citep{Bower:2001}
have proven to be useful summary pictures of such a biochemical system or part
of it, especially when interactions are known qualitatively while precise
quantitative information is scarce. For many systems, the interaction network
suffices to capture essential features of dynamical behaviour \citep{Albert:2003}
such as the presence of a stable stationary state, multistability, oscillations
etc. Often such predictions do not even depend on the whole network
structure. Qualitative statements on system behaviour may be based on the
centrality of nodes \citep{Jeong:2001,Wuchty:2003} or the presence of
certain small subnetworks, called motifs \citep{Alon:2007}.

Here we ask to what extent a limited knowledge of biochemical interactions is
usable for predicting the reaction of a system to failure of one or several of
its components \citep{Albert:2000,Inger:2009,Boldhaus2010b}. This kind of
theory serves to complement experiments with knock-outs in vivo or in vitro
\citep{deVisser:2003}. A knock-out (or knock-down) is performed by blocking (or
reducing) production of a single protein. Depending on the viability of the
cell after suffering the modification, knock-outs are subject to a binary
classification into {\em lethal} and {\em viable}. 

When knocking out several nodes (proteins) of a system simultaneously, a richer
classification arises from considering the lethality of this combined knock-out
together with the effect of each single knock-out. {\em Synthetic lethality}
\citep{Hartman:2001} is the class of lethal simultaneous knock-out of two nodes $i$
and $j$, where independent knock-out of node $i$ alone is viable and independent
knock-out of node $j$ alone is viable. Synthetic lethality has direct
implications for target identification in anticancer drug discovery
\citep{Chan:2011}. Since the experimental effort of a complete scan of double
knock-outs is quadratic in the number of proteins, an accurate computational
prediction of candidate pairs can greatly reduce the cost of experiments.

Here we study prediction of lethality and synthetic lethality in stochastically
generated interaction networks. Knowledge is taken to be incomplete in the sense
that only the absence or presence of interactions but not the type (enhancer /
suppressor) is available. As predictors, we test efficiently computable
network centrality measures based on degree and betweenness. Quality of
predictors in terms of ROC curves (see Section~\ref{sec:Methods}) is held against the optimal
prediction possible with the available knowledge. Additionally, we use
evolutionary distance between nodes as a predictor. Thereby we find out how
much the knowledge of paralogs supports the identification of synthetically
lethal pairs.

Our notion of viability and lethality is based on a functional phenotype that we
define here as a stationary state of the unperturbed dynamical system.
Regulatory interactions are mimicked by Boolean threshold dynamics that serves
as a suitable testbed for the studies of robustness of networked biological
systems \citep{Bornholdt2005} and for evolutionary studies \citep{Luo:2011}.

\section{Network construction} \label{sec:Networks}

Throughout this contribution, a {\em network}\ on $n$ nodes is given by an
$n \times n$ matrix $W$. Each matrix entry $w_{ij}$ takes a value in
$\{-1,0,+1\}$ where
\begin{equation} 
w_{ij}= \left\{ \begin{array}{rl}
+1, & \text{if $j$ is an enhancer of $i$} \\
-1, & \text{if $j$ is a suppressor of $i$} \\
0,  & \text{otherwise} \end{array} \right. ~.
\end{equation}
Reduced information about interactions is represented by assigning the network $W$
a directed graph that we identify with its adjacency matrix $A$. The entries of
$A$ are given by $a_{ji} = | w_{ij} |$. Thus $A$ contains the information about
the absence or presence but not the type of a directed interaction. The
density $\rho(W)$ of a network 
\begin{equation}
\rho(W) = n^{-2} \sum_{i=1}^n \sum_{j=1}^n |w_{ij}|
\end{equation}
measures the fraction of interactions established out of the $n^2$ possible
ones. We generate networks with the following two stochastic procedures.

\subsection{Random networks}
A random network $W$ is generated by independently assigning each entry
$w_{ij}$ a value $+1$ with probability $p/2$, a value $-1$ with probability
$p/2$ and a value $0$ with probability $1-p$. The model parameter $p$ is to
be chosen from $[0,1]$ and determines the average density of the random
network \citep{Drossel2007,Aldana2003b}.

\subsection{Networks from duplication and divergence} \label{sec:DaD}
\begin{figure}
\centerline{\includegraphics[width=\linewidth]{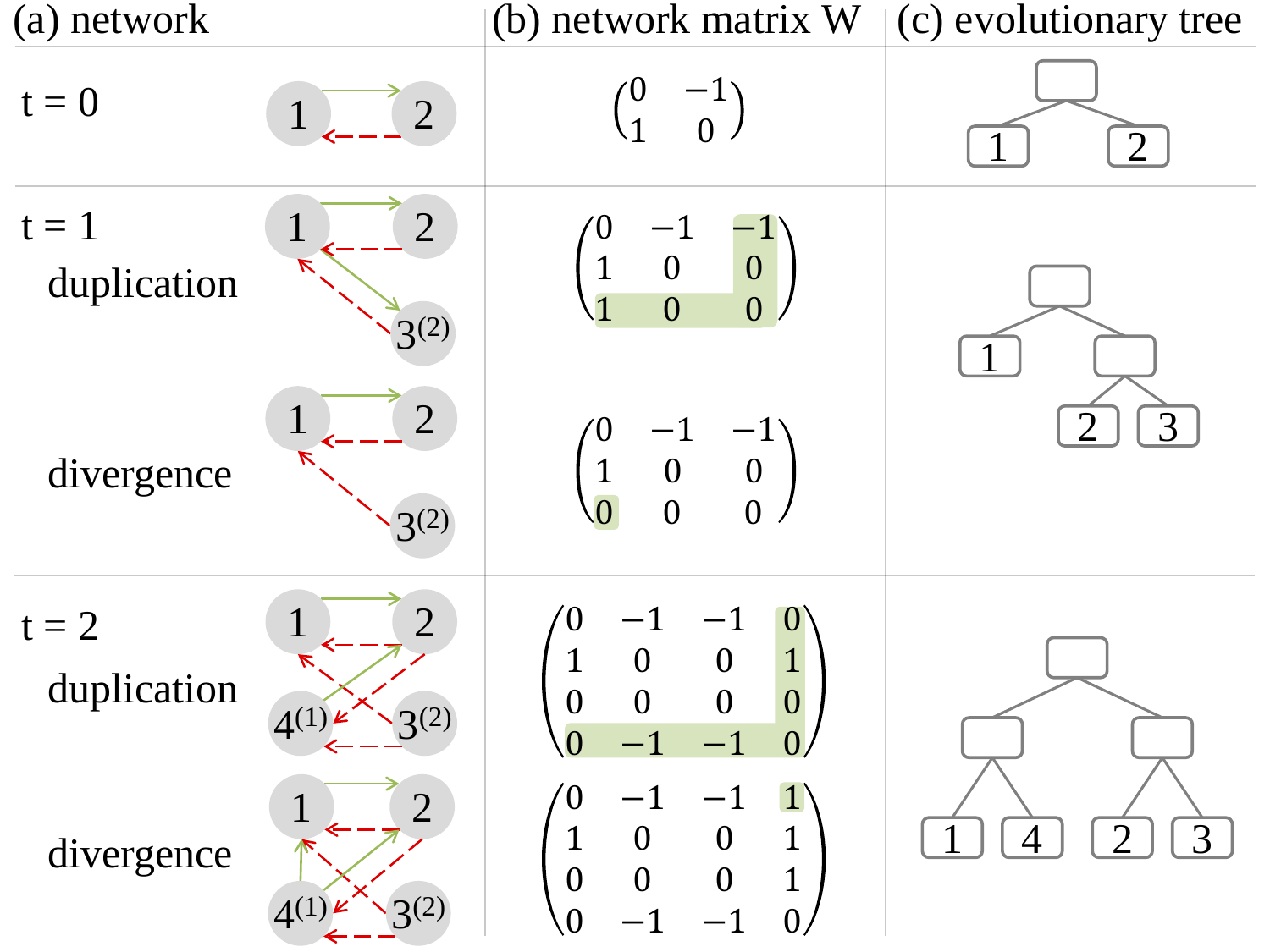}}
\caption{\label{fig:GenerateNet}
Example of network generation of a network with duplication and
divergence. (a) The network at time-step $t=0$ is initialized with two
asymmetrically coupled nodes. Afterwards it grows by successive duplication
and divergence steps.
(b) The growth process in terms of the network
matrix $W$. (c) In the evolutionary tree, each leaf represents an extant
node in the network. Inner nodes are common ancestors.}
\end{figure}

An alternative statistical ensemble of networks is generated by duplication
and divergence. This is motivated by the observation that an evolutionary
extension of the repertoire of regulatory sequences is obtained by
duplication 
\citep{Wagner1994,Sole2002,PastorSatorras:2003,Ispolatov2005a,Aldana:2007}.

For generating a network by duplication and divergence (DaD) we start with
a $2 \times 2$ matrix representing two mutually but unequally coupled
nodes, i.e.{} $w_{12} = +1$, $w_{21} = -1$ and $w_{11} = w_{22} = 0$. Then the
following process of duplication (i) and divergence (ii) is iterated.

\begin{itemize}
\item[(i)] A node~$i$ of the network with $n-1$ nodes is randomly drawn
from a flat probability distribution. Node~$i$ is duplicated, generating an
additional row and column with index $n$ in the matrix $W$.  The new
entries are $w_{j,n} := w_{j,i}$ and $w_{n,j} := w_{i,j}$ for all $1 \leq j
< n$, and $w_{n,n}=w_{i,i}$.
\item[(ii)] For each index pair $(k,l)$ with $k=n$ or $l=n$: if $|w_{kl}|=1$,
we set $w_{kl}:=0$ with probability $r$ and leave $w_{kl}$ unchanged with
probability $1-r$. Otherwise ($w_{kl}=0$), we set $w_{kl}:=+1$ with probability
$a/2$, $w_{kl}:=-1$ with probability $a/2$ and leave $w_{kl}$ unchanged with
probability $1-a$.
\end{itemize}

Step (i) implements gene duplication, in which both the original and the
replicated proteins  retain the same structural properties and the same set of
interactions. The divergence steps (ii) implements the possible mutations of
the replicated gene, which translate into the addition and removal
of interactions with probabilities $a$ and $r$. An example of this
process is shown in Figure~\ref{fig:GenerateNet}. Special attention is given to
the handling of loops. If the randomly chosen original node has a loop, the
loop is copied as well as two additional links with the same edge weight
between the original and the replica node.

\begin{figure}
\centerline{\includegraphics[width=\linewidth]{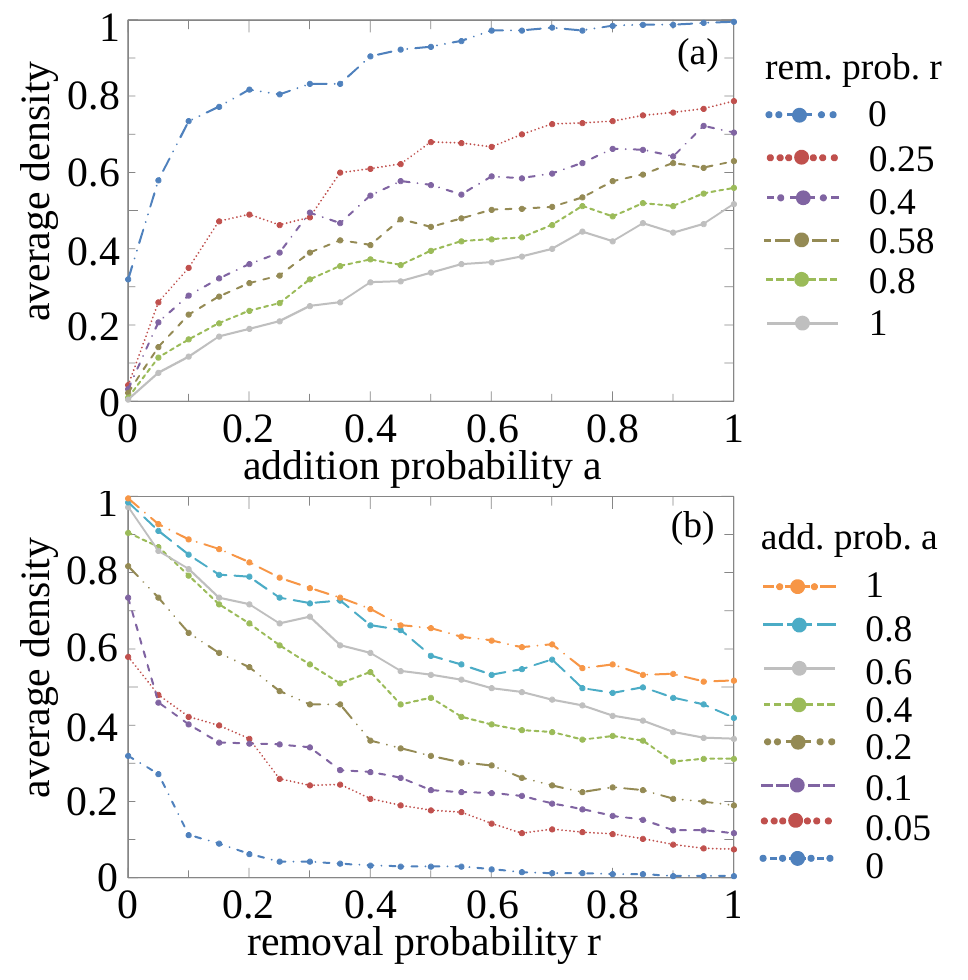}}
\caption{Average densities for networks generated with duplication and
divergence. (a) Average density with fixed removal probabilities~$r$. (b)
Average density with fixed addition probabilities~$a$. All points are
averages over $10^6$ realizations \revision{of networks with $n=20$ nodes.
}} \label{fig:averageDegree}
\end{figure}
The average density of networks generated with duplication and divergence is
shown in Figure~\ref{fig:averageDegree} as a function of parameters $a$ and $r$. 

\section{Knock-outs, dynamics, and functionality} \label{sec:dyn}
\subsection{Knock-outs} \label{subsec:knock}
In a real biochemical interaction network, knocking out a node means that the
concentration of the reactant represented by the node is set
zero. For our purposes, it is equivalent to remove all outgoing interactions
(arcs) of the node from the network. The knock-out of node $k$ in network $W$
leaves the network as $W^{\setminus\{k\}}$ with matrix entries
$w_{ij}^{\setminus\{k\}} = w_{ij}$ if $j \neq k$ and $0$ otherwise. As a
generalization, several nodes forming a set {$K \subset \{1,2,\dots,n\}$}
may be knocked out. The resulting network $W^{\setminus K}$ has entries 
\begin{equation}
w_{ij}^{\setminus K} = \left\{ \begin{array}{ll}
w_{ij} & \text{if } j \notin K\\
0      & \text{otherwise.}
\end{array} \right. ~. 
\end{equation}

\subsection{Dynamics}
In the present work we model gene regulatory networks by threshold dynamics
\citep{Derrida1987}.  This is a special case of Boolean dynamics
\citep{Kauffman1969,Aldana2003b}.  A multitude of formulations for Boolean
threshold networks exist, depending how the thresholds are distributed, how
the behaviour at the threshold is defined and which weights for the edges
are allowed. We choose the version as applied in the simplified yeast cell
cycle network \citep{Li:2004} and many succeeding studies
\citep{Boldhaus2010a,Boldhaus2010b,Szejka2004,Davidich2008a}, compare
Equation~\ref{eq:dyn}.

A node is activated, $s_i = 1$, if the sum of its weighted inputs exceeds a
threshold assumed to be zero here. It is deactivated if the input
sum falls below the threshold. In the case when the sum gives exactly the
threshold value, the node value remains unchanged. Thus the Boolean state
$s_i$ of node $i$ at time step $t$ evaluates to
\begin{equation} \label{eq:dyn}
s_{i}(t+1)=\left\{
\begin{aligned}
1 \textrm{ if }
    & \textstyle\sum_{j=1}^n w_{ij} s_{j}(t) > 0 \\
0 \textrm{ if }
    & \textstyle\sum_{j=1}^n w_{ij} s_{j}(t) < 0 \\ 
s_i(t) \textrm{ if }
    & \textstyle\sum_{j=1}^n w_{ij} s_{j}(t) = 0
\end{aligned} 
\right. ~.
\end{equation}

\revision{This threshold dynamics does not capture the wealth of combinatorial
effects implementable by control at the transcriptional level
\citep{Buchler:2003}. However, it is able to account both for cooperative
and non-cooperative interactions. In a network with $w_{ij}=w_{ik}=1$ being the
only incoming connections of node $i$, for instance, these two 
inputs $j$ and $k$ act non-cooperatively on node $i$. Then
$s_j(t)=1 \vee s_k(t)=1$ is sufficient for $s_i(t+1)=1$, amounting
to an {\sc or} operation.}

\subsection{Functionality and lethality}
A state $s^\ast$ is a {\em fixed point} if it remains unaltered by the
dynamics, i.e.\ the successor state of $s^\ast$ is $s^\ast$ itself. We define the
{\em function} (in the sense of a phenotype) of a network $W$ to be a fixed
point state $s^\ast \neq (0,\dots,0)$.  After generation of a network, a fixed
point $s^\ast$ is found as described in Section~\ref{sec:Methods}.

Given a network $W$ and its functional fixed point $s^\ast$, we say that a
knock-out $K \subset \{1,\dots,n\}$ is {\em viable}\ (for $W,s^\ast$) 
if $s^\ast$ is a fixed point
of $W^{\setminus K}$. Otherwise $K$ is {\em lethal}. \revision{Note that
a sufficient (but not necessary) condition for $K$ to be viable
is that $s_i^\ast=0$ for all $i \in K$: a lethal effect is not caused by
knocking out nodes that are inactive already.} 

We say that $K$ is {\em synthetically lethal}, if
\begin{itemize}
\item[(i)]  $K$ is lethal and 
\item[(ii)] $K^\prime$ is viable for all proper subsets $K^\prime \subset K$. 
\end{itemize}
Analogously one may define {\em synthetic viability}.
A knock-out $K$ is synthetically viable, if
\begin{itemize}
\item[(i)]  $K$ is viable and 
\item[(ii)] $K^\prime$ is lethal for all proper subsets $K^\prime \subset K$,
$K^\prime \neq \emptyset$. 
\end{itemize}
Thus synthetic lethality and synthetic viability are defined for arbitrary
cardinality $|K|\ge 2$ of knock-outs. In this contribution, however,
only single and double knock-outs are considered.

\revision{One might wonder in which sense our definitions match a possibly more
intuitive definition of a knock-out. Naturally, a knock-out of a node $i$
could be defined as a modification of the dynamical rules, Eq.~(\ref{eq:dyn}),
where we keep $s_i$ at value zero, irrespective of the input signals node $i$
receives. Then for such a knock-out to be called viable, we would require that
this modified dynamics has a fixed point $r^\ast$ with $r_j^\ast=s_j^\ast$ for
$j \neq i$ and $r_i^\ast=0$ otherwise. Let us compare this to the above
definitions. Rather than changing the dynamical rules, the network itself is
modified by removing all outgoing interactions of the node knocked out. On all
other nodes, this has the same effect as keeping $s_i$ at state zero. Then, if
the dynamics has a fixed point that coincides with $s^\ast$ on all nodes $j \neq
i$, also node $i$ will be in state $s_i^\ast$ at this fixed point. Hence the
present definitions, while being convenient and concise in notation, coincide
with the intuitive notion.}

\section{Results} \label{sec:Results}

\revision{All results presented in this section are based on simulations
with networks having $n=20$ nodes. The Supplementary Material provides 
results for larger and smaller networks for comparison.}

\begin{figure}
\centering
\centerline{\includegraphics[width=\linewidth]{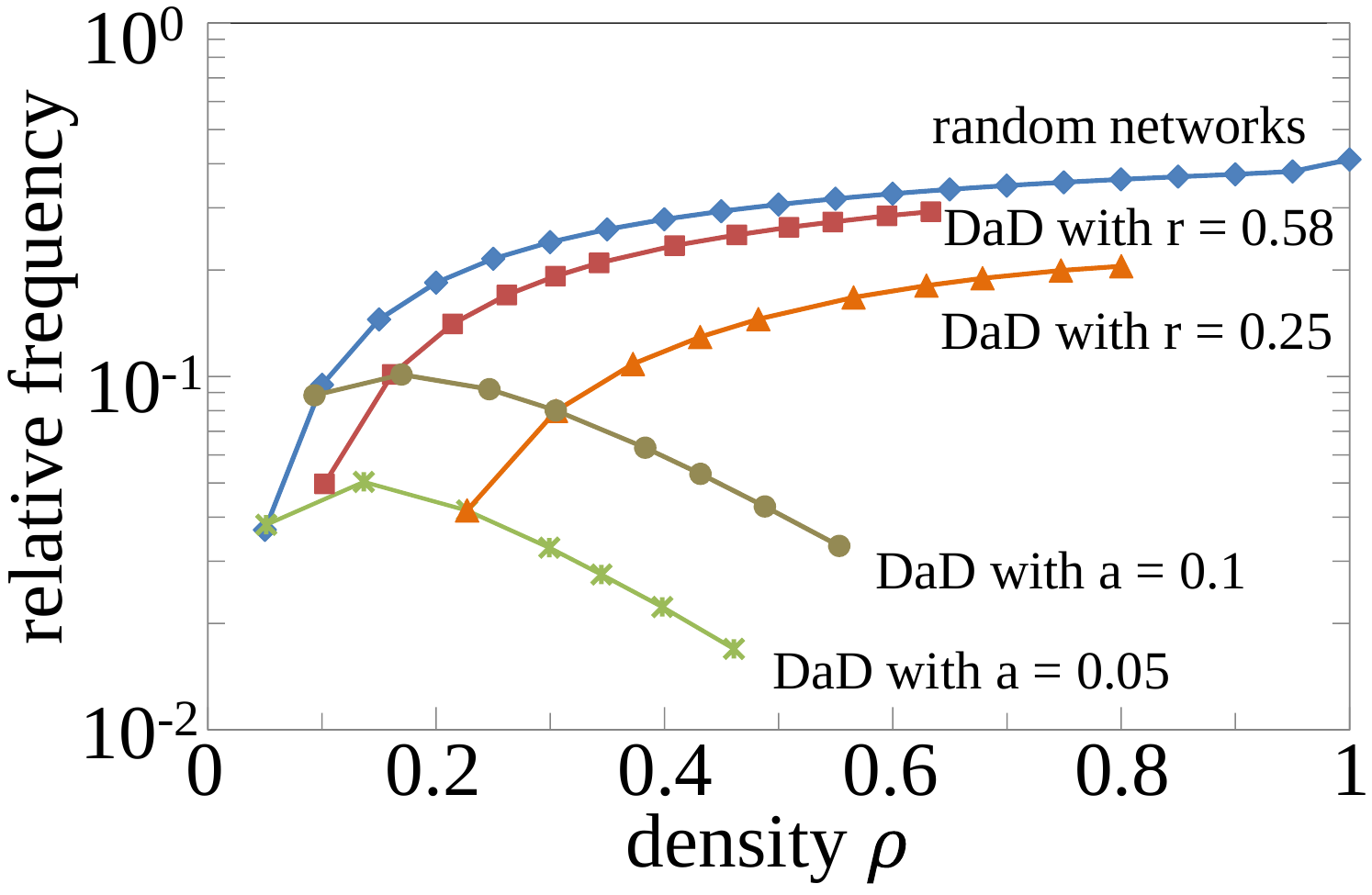}}
\vspace{-0.3cm}
\caption{\label{fig:frac_single}
Probability of lethal single node knock-outs as a function of network density $\rho$.
All values are averages over $10^6$~realizations at the given value of $\rho$.
Holding the addition (removal) probability~$a$ ($r$) constant limits the
interval of possible densities.}
\end{figure}

\begin{figure}
\centerline{\includegraphics[width=\linewidth]{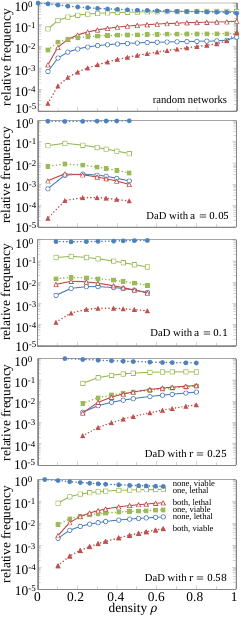}}
\caption{\label{fig:scenarios}
Lethality of knock-outs as a function of network density~$\rho$. 
Open symbols refer to lethal double knock-outs while solid
symbols mean that the double knock-out is still viable. The shape of the
symbols distinguishes between results of single-node knock-outs: neither
single knock-out (circle), only one knock-out (square) or both single knock-outs are lethal (triangle). 
All values are averages over $10^6$~realizations at the given value of $\rho$.}
\end{figure}

\begin{table*}
\caption{\label{tab:singleKO}
Overview of the area under the ROC curves for prediction of single
node knock-outs. Each pair of rows is for networks with a given expected
density $p$. The first row of each pair refers to random networks with
parameter value $p$. The second row of each pair is for DaD networks with parameter values
$a$ and $r$. Higher values of the area refer to a higher accuracy of
the prediction.}
\begin{tabular*}{\textwidth}{lccccccc}
\hline
	\multicolumn{1}{c}{} &
	\multicolumn{1}{c}{struct.\ lethality} &
	\multicolumn{1}{c}{out-deg.} &
	\multicolumn{1}{c}{out$~+~$in-deg.} &
	\multicolumn{1}{c}{out$~-~$in-deg.} &
	\multicolumn{1}{c}{betw.{}centr.} &
	\multicolumn{1}{c}{in-deg.}\\
\hline \hline
$p=0.14$ 						& 0.672 & 0.649 & 0.621 & 0.591 & 0.566 & 0.483\\ 
$a=0.05$,~$r=0.58$ 	& 0.732 & 0.708 & 0.634 & 0.662 & 0.593 & 0.486\vspace*{2mm}\\ 
$p=0.23$ 						& 0.623 & 0.612 & 0.587 & 0.572 & 0.560 & 0.492\\ 
$a=0.1$,~$r=0.58$ 		& 0.681 & 0.666 & 0.608 & 0.626 & 0.577 & 0.494\vspace*{2mm}\\ 
$p=0.26$						& 0.612 & 0.602 & 0.579 & 0.566 & 0.557 & 0.494\\ 
$a=0.05$,~$r=0.25$ 	& 0.689 & 0.682 & 0.635 & 0.617 & 0.598 & 0.534\vspace*{2mm} \\ 
$p=0.35$ 						& 0.588 & 0.581 & 0.562 & 0.553 & 0.548 & 0.496\\ 
$a=0.1$,~$r=0.25$ 		& 0.648 & 0.642 & 0.602 & 0.596 & 0.574 & 0.518\\ 
\hline \end{tabular*}
\end{table*}

\begin{figure}
\centerline{\includegraphics[width=\linewidth]{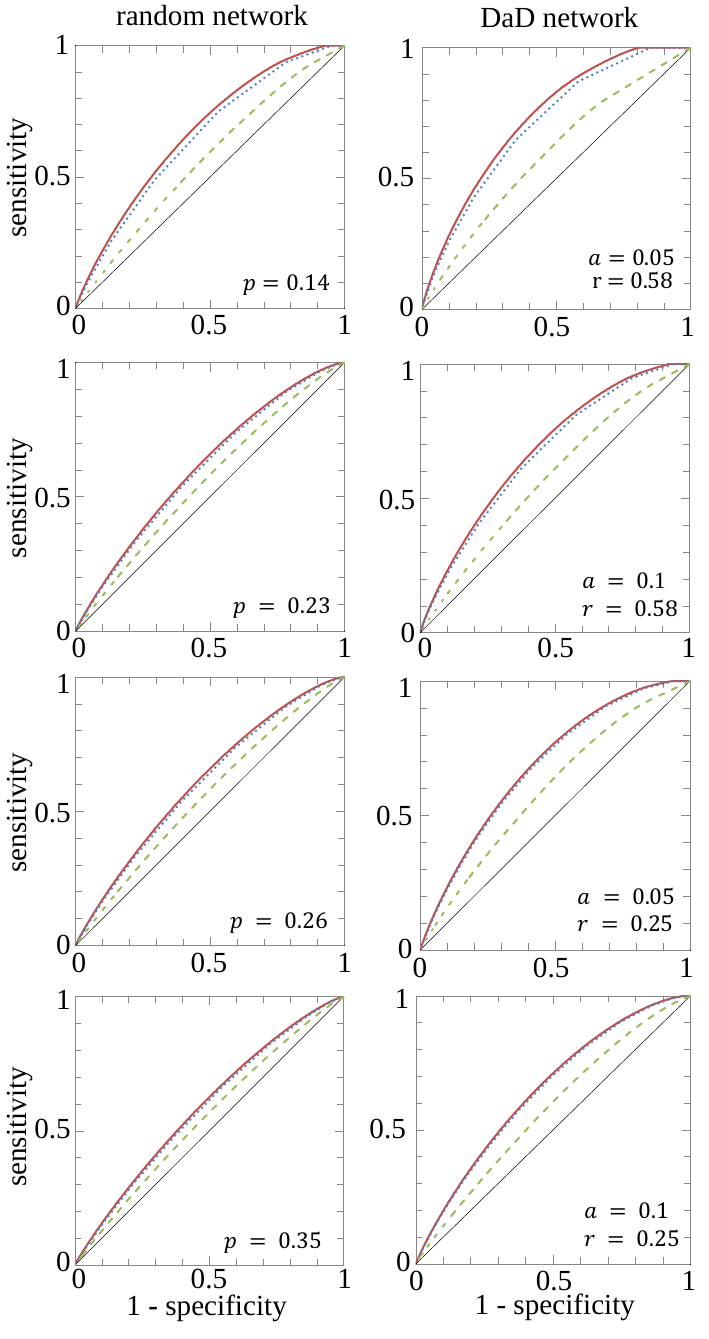}}
\caption{\label{fig:roc_single}
ROC curves for prediction of lethality of single node knock-outs. Average
network density is $\rho = p$ for the random networks (panels in left column).
In the corresponding panel in the right column,  results for DaD networks with
the same density are shown. Predictors are structural lethality (solid curve,
optimal predictor), out-degree (dotted), and betweenness centrality (dashed).
The solid diagonal is the line of no discrimination. Each curve is based on
$10^4$ network realizations. The  synthetic lethality is estimated by
extracting the graph from each network and probing another $10^3$ network
realizations with the same graph structure.
} 
\end{figure}

\begin{table*}
\caption{\label{tab:doubleKO}
Overview of the area under the ROC curves for prediction of double node
knock-outs which exhibit synthetic lethality. Each pair of rows is for networks
with a given expected density $p$. The first row of each pair refers to random
networks with parameter value $p$. The second row of each pair is for DaD
networks with parameter values $a$ and $r$. Prediction based on
evolutionary distance is only applicable for networks generated with
duplication and divergence. Results which incorporate prior knowledge of
the result of single node knock-outs are shown in brackets. Higher values
of the area refer to a higher accuracy of the
prediction.}
\begin{tabular*}{\textwidth}{lccccc}
\hline 
	\multicolumn{1}{c}{} &
	\multicolumn{1}{c}{struct.\ syn.\ let.} &
	\multicolumn{1}{c}{out-overlap} &
	\multicolumn{1}{c}{repl. centr.} &
	\multicolumn{1}{c}{evol. distance} &
	\multicolumn{1}{c}{in-overlap}\\
\hline  \hline 
$p=0.14$ & 0.888 (0.895) & 0.859 (0.865) & 0.597 (0.600) & - & 0.500 (0.499)\\ 
$a=0.05$,~$r=0.58$ & 0.915 (0.922) & 0.896 (0.903) & 0.594 (0.597) & 0.601 (0.601) & 0.531 (0.530)\vspace*{2mm}\\ 
$p=0.23$ & 0.778 (0.787) & 0.742 (0.752) & 0.582 (0.586) & - & 0.501 (0.500)\\ 
$a=0.1$,~$r=0.58$ & 0.857 (0.867) & 0.832 (0.841) & 0.588 (0.591) & 0.550 (0.551) & 0.519 (0.519)\vspace*{2mm}\\ 
$p=0.26$ & 0.743 (0.752) & 0.705 (0.717) & 0.576 (0.581) & - & 0.500 (0.500)\\ 
$a=0.05$,~$r=0.25$ & 0.799 (0.812) & 0.779 (0.789) & 0.605 (0.609) & 0.613 (0.611) & 0.573 (0.572)\vspace*{2mm}\\ 
$p=0.35$ &  0.673 (0.681) & 0.632 (0.646) & 0.559 (0.563) & - & 0.500 (0.500)\\ 
$a=0.1$,~$r=0.25$ & 0.735 (0.748) & 0.707 (0.724) & 0.583 (0.587) & 0.570 (0.568) & 0.546 (0.546)\\ 
\hline
\end{tabular*}
\end{table*}

\begin{figure}
\centerline{\includegraphics[width=\linewidth]{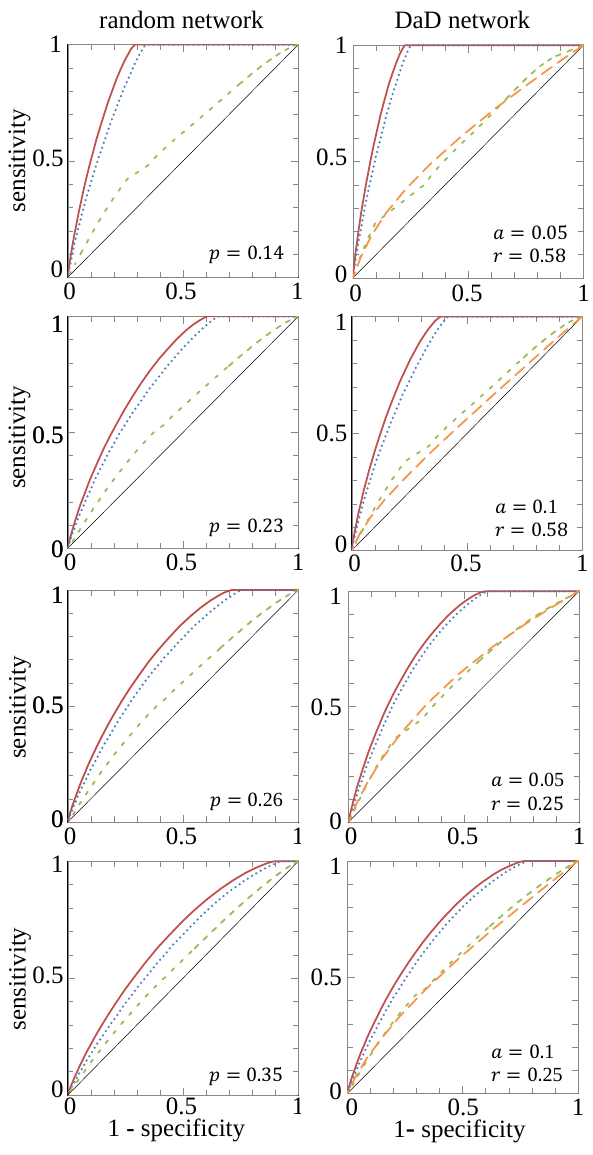}}
\caption{\label{fig:roc_double}
ROC curves for prediction of synthetic lethality. Average network density is
$\rho = p$ for the random networks (panels in left column). In the
corresponding panel in the right column,  results for DaD networks with the
same density are shown. Predictors are structural synthetic lethality (solid
curve, optimal predictor), out-overlap (dotted), replacement centrality
(short-dashed), and evolutionary distance (long-dashed). The solid diagonal is
the line of no discrimination. Each curve is based on $10^4$ network
realizations. The structural synthetic lethality is estimated by extracting
the graph from each network and probing another $10^3$ network realizations
with the same graph structure.
} 
\end{figure}

\subsection{Statistics of lethal knock-outs}

We start by presenting the effect of single knock-outs in the two models of
networks (random and DaD). Figure~\ref{fig:frac_single} shows, as a function of
the density of the network, the probability that a single knock-out is lethal.
For random networks, this probability increases with the arc density
$\rho$. For all
choices of parameters investigated here, DaD networks have fewer lethal
knock-outs than random networks at the same density. Under constant probability
$a$ of adding interactions, the DaD networks even exhibit a decreasing
number of lethal knock-outs with increasing density.

Note that the probability of knock-outs being lethal cannot exceed $1/2$ because
on average half of the nodes are in the off-state on the functional fixed point.
Knock-out of a node in the off-state in the network does not affect the states
of the other nodes. This theoretical maximum, however, is not reached. Random
networks at the largest possible density 1 have a probability of $\approx 0.41$ for 
a knock-out to be lethal. 

Now we turn to the statistics for double knock-outs $\{v,w\}$ in
combination with the single knock-outs $\{v\}$ and $\{w\}$. In each of the
panels of Figure~\ref{fig:scenarios}, the open circles connected by solid
curves give the fraction of synthetically lethal pairs of nodes in networks
of a given density $\rho$. Synthetic lethality becomes more abundant with
increasing density in random networks (top panel) and in DaD networks with
fixed arc removal probability (two lower panels).

Synthetic lethality is just one possible outcome of knock-out tests performed on
a pair $\{v,w\}$. Of the single-node knock-outs $\{v\}$ and $\{w\}$, none,
exactly one or both may be lethal. Combination of this ternary result with  
the binary outcome (lethal/ viable) of the two-node knock-out
$\{v,w\}$ yields six possible scenarios. The statistics
of these scenarios is plotted in Figure \ref{fig:scenarios}. Interestingly,
synthetic viability (dotted line with triangles) becomes more abundant
than synthetic lethality in dense random networks.

\subsection{Prediction} \label{sec:prediction}


Now we investigate the predictability of the lethality, first of single,
then of double knock-outs. Predictability is strongly dependent on the
available knowledge. If the network $W$ and the functional fixed point
$s^\ast$ are fully known, perfect prediction of lethality is possible simply
by computing the effect of the knock-out. In a more realistic scenario,
only partial knowledge is available which we model here as follows.
The presence or absence of each interaction is available
with absolute accuracy while the type  (enhancer/ suppressor) of each present
interaction is totally unknown; information on the functional fixed point
is not available. With the formalism described in
section~\ref{sec:Networks}, for a network $W$ only the adjacency matrix $A$ of
the directed graph is known. Then the best predictor of lethality of a
knock-out $\{i\}$ is the relative frequency $l_i$ of $\{i\}$ being lethal
in networks $W^\prime$ that have adjacency matrix $A$ and interaction types
randomly assigned. In other words, $\{i\}$ is predicted as lethal in $W$,
if $\{i\}$ is typically lethal in networks with the same adjacency matrix
as $W$. We call $l_i$ structural lethality.

Computation of $l_i$ may be impossible or impractical in real scenarios.
Therefore the value of a centrality measure at node $i$ is extracted from $A$
and used for prediction instead of $l_i$. This incurs another step of knowledge
reduction. Here it is salient to choose the ``right'' centrality
measure for prediction. We consider
the betweenness centrality $b_i$, the out-degree $d_i^\text{out}$
the in-degree $d_i^\text{in}$, furthermore the degree sum
$d_i^\text{out}+d_i^\text{in}$ and the degree difference
$d_i^\text{out}-d_i^\text{in}$.
These are defined in Section~\ref{sec:Methods} as well as other quantities
used here.  The predictive power of the different measures is summarized in
Table~\ref{tab:singleKO} for random and DaD models with varied parameter
values.
The deviation of a value from $0.5$ indicates that the quantity contains
information about the lethality of nodes in the given scenario. This is
the case for all the centrality measures under consideration except for the
in-degree. In DaD networks the chance to predict the effect of a knock-out is larger
than in random networks of the same density, e.g.\,$a=0.1$, $r=0.25$ leads to
networks with density $p \approx 0.35$ which show a larger
predictability than random networks with the same density.
Best predictions are based on the out-degree whose predictive
power almost reaches the best possible obtained by the structural
lethality. Predictive power is measured as the area under the ROC curve of sensitivity
versus specificity of the prediction which are shown in
Figure~\ref{fig:roc_single}.


Now we study the prediction of synthetically lethal pairs. The framework
is mostly analogous to that of single knock-outs. Again we assume that
the adjacency matrix but not the full network (with interaction types)
are known. Eligible predictors are now measures of {\em joint} centrality,
i.e. mapping a given unordered node pair $\{i,j\}$ in a given graph $A$ to
a number. Here we investigate the out-overlap $o^\text{out}_{ij}$,
the in-overlap $o^\text{in}_{ij}$ and the replacement centrality
$r_{ij}$ as defined in section~\ref{sec:Methods}. Furthermore, the
evolutionary distance $e_{ij}$ is used as a predictor in networks
evolved by duplication and divergence (DaD).
Table~\ref{tab:doubleKO} summarizes the predictive power of these centrality
measures, again in comparison with that of the optimal predictor $s_{ij}$
here called {\em structural synthetic lethality}. As the main result, the
out-overlap $o^\text{out}_{ij}$ is the best predictor of synthetic lethality
in all cases considered, its predictive power is close to optimal in all
cases considered. Prediction of synthetic lethality is facilitated in DaD
networks as compared to random networks.

An interesting alternative scenario arises under the assumption that we already
know all viable single node knock-outs, $V(W)= \{i : \{i\} \text{ viable} \}$ in
each network $W$ considered. For the prediction of synthetically lethal pairs,
candidates are subsets $\{i,j\} \subseteq V(W)$, $i \neq j$. The predictive
power for this scenario with restricted candidate set is given in brackets in
Table~\ref{tab:doubleKO}. Prior knowledge of viable single-node knock-outs does
not induce a significant increase of predictive power for any of the
combinations of predictor, network generation model and density.

For a more detailed view of sensitivity and specificity,
Figure~\ref{fig:roc_double} shows the ROC curves of selected predictors in the 
case without prior knowledge of viable single knock-outs.

\section{Concluding remarks} \label{sec:Concl}

The present contribution has established a theoretical framework for assessing
predictability of knock-out effects in networked regulatory systems.
We covered a broad range of scenarios in terms of network structures and
predictors. \revision{Results are robust under variation of system size and choice
of functional fixed points, cf.\ additional results in Supplementary Material.}

Nevertheless it must be emphasized that outcomes depend on the choice of
specific definitions made. First of all, the definition of lethality is made in
the context of the regulatory network, assuming that the only task of a node is
regulation within the system considered. If the disabled protein is involved
otherwise, e.g.{} as a structural protein, its potentially lethal knock-out
cannot be predicted in the present framework. Secondly, the definition of
functionality of the network as the presence of a fixed point is not the only
reasonable choice. For instance, we may demand that the fixed point $s^\ast$ be
stable in the sense that the system returns to $s^\ast$ after a perturbation at
one node's state. Rather than a fixed point, a particular temporal sequence of
states may be defined as the functionality of the system
\citep{Li:2004,Davidich2008a,Boldhaus2010b,Luo:2011}.

\revision{The aim of the present study is to contribute to the theoretical
background of knock-out experiments. It elucidates in how far the lethal
effect of knock-outs is predictable by efficiently computable measures of
node centrality. A future extension may be concerned with the effect of disabling or
modifying single regulatory interactions rather than entirely knocking out
genes. Such a scenario corresponds to natural or experimentally induced
mutations of transcription factor binding sites.  
Analogous to the present study, a comparison of measures of {\em edge}
centrality can find the best predictors for the lethality of such mutations. 
Alternatively, the present scenario using node centralities may be applied to
the line graph of the regulatory network.}

\section{Methods} \label{sec:Methods}
\subsection{Finding functional fixed point}
After generating a network $W$, the functional fixed point $s^\ast$ is assigned
as follows. An initial state vector $s(0) \in \{0,1\}^n$ is drawn uniformly. The
dynamics is run from $s(0)$ by iterating Equation~(\ref{eq:dyn}) until a state is
seen the second time and an attractor is reached, i.e.{} at times $t_1>t_2\ge
0$ such that $s(t_1) = s(t_2)$. If the attractor is a non-trivial fixed
point, $s(t_2)=s(t_2-1)\neq (0,0,\dots,0)$, we take it as the functional
fixed point $s^\ast :=s(t_2)$.  Otherwise the network $W$ is discarded and
replaced by another random instance. \revision{This procedure preferentially
chooses functional fixed points with larger attractor basins.

For comparison, the Supplementary Material provides additional
results obtained by a different procedure for choosing the functional fixed
point. We first determine the set $F \subseteq \{0,1\}^n \setminus\{0,\dots,0)\}$
of fixed points of the given network $W$.
If $F$ is not empty, the functional fixed point $s^\ast$ is drawn uniformly
from $F$. Otherwise the network $W$ is discarded and
replaced by another random instance.}

\subsection{Evolutionary distance} \label{DefEvolDist}
Along with the generation of a network $W$ with the DaD model (Section~\ref{sec:DaD}),
the evolutionary tree $T$ is constructed, cf.\ the example in
Figure~\ref{fig:GenerateNet}. The nodes $\{1,2,\dots,n\}$ of the network $W$ are
the leaves of the tree $T$. The evolutionary distance $e_{ij}$
is defined as the length of the path between leaves $i$ and $j$ on $T$.
Each edge on $T$ is counted with unit length.

\subsection{Measures of centrality}

\paragraph{Degree centralities} They measure importance in a linear fashion,
assuming that a node with twice the number of links also is twice as
important. Several different degree centralities can be defined by using a 
function of the in-degree~$d_i^{\rm in}$ and the out-degree~$d_i^{\rm out}$
of a node~$i$ for a given adjacency matrix $A$:
\begin{equation}
d_i^\text{in}(A) = \sum_{j=1}^n a_{ji}~ \qquad
d_i^\text{out}(A) = \sum_{j=1}^n a_{ij}~.
\end{equation}

\paragraph{Overlaps} We distinguish the in-overlap $o^\text{in}$
between nodes $i$ and $j$ as
\begin{equation}
o^\text{in}_{ij}(A) =  \frac{| \{ k : a_{ki}=1 \wedge a_{kj}=1 \} | } 
                 {| \{ k : a_{ki}=1 \vee a_{kj}=1 \} |} 
\end{equation}
and the out-overlap $o^\text{out}$
\begin{equation}
o^\text{out}_{ij}(A) = \frac{| \{ k : a_{ik}=1 \wedge a_{jk}=1 \} | } 
                 {| \{ k : a_{ik}=1 \vee a_{jk}=1 \} |} 
\end{equation}
If the denominator is zero, the whole expression is defined to be zero.

\paragraph{Betweenness centrality} 
It quantifies the fraction
of shortest paths that pass through this node \citep{Freeman1977}.
\begin{equation}
b_i (A) = \sum_{(j,k)} \frac{\sigma_{jk}(i)}{\sigma_{jk}}~,
\end{equation}
where the sum runs over all ordered node pairs $(j,k)$; $\sigma_{jk}$
denotes the total number of shortest paths from node $j$ to node $k$;
$\sigma_{jk}(i)$ is the number of such paths running through node $i$.
A modified Floyd-Warshall algorithm \citep{Brandes2001} allows to
simultaneously compute the lengths and numbers of shortest paths.

\paragraph{Replacement centrality} 
Let us define the {\em replacement centrality} of a pair of nodes $(i,j)$ as
\begin{equation} \label{eq:repcent}
r_{ij} = \frac{b_i (A^{\setminus\{j\}}) + b_j (A^{\setminus\{i\}})}{b_i(A) + b_j(A)}
\end{equation}
if the denominator in Equation~(\ref{eq:repcent}) is non-zero and $r_{ij}(A)=0$ otherwise. 

\paragraph{Structural synthetic lethality} 
When only knowing the graph $A$, the best predictor of synthetic lethality
for nodes $i$ and $j$ is given by the fraction of networks exhibiting
synthetic lethality at $i$ and $j$ out of all networks compatible with $A$.
This fraction is called structural synthetic lethality and is formally defined
as
\begin{equation}
s_{ij}(A) = \frac{ | \{W \in {\cal N}(A) | \{i,j\} \text{ synth.\ lethal in }W \} | }
                 { | {\cal N}(A) |}
\end{equation}
where ${\cal N}(A)$ is the set of all networks that map to the graph $A$.  

\subsection{Receiver Operating Characteristic (ROC)} \label{DefRoc} 
The \emph{sensitivity} of a prediction is the fraction of cases
for which the outcome is positive and correctly identified.
Similarly, the \emph{specificity} is the fraction of cases correctly
identified as negative. 
A Receiver Operating Characteristic (ROC) \citep{Fawcett2006} is the collection of all tuples of
(specificity, sensitivity) obtained by varying a threshold $\theta$ 
on the quantity used as a predictor.

Formally, we consider a set of objects $S$ with a binary partition, i.e.{} subsets $S^+$ and $S^-$
with $S^+ \cap S^- = \emptyset$ and $S^+ \cup S^- = S$. As a predictor of this
partition, we consider a
function $v: S \rightarrow \mathbb{R}$ and a threshold value $\theta \in \mathbb{R}$.
An object $x \in S$ is predicted as positive ($+$) if $v(x) \ge \theta$.
The sensitivity measures the fraction of objects from $S^+$ predicted to be
positive ($+$), i.e.{}  
\begin{equation}
\frac{|\{x \in S^+ : v(x)\ge \theta\}|} {|S^+|}~.
\end{equation}
and analogously, the specificity is the fraction of objects from $S^-$
predicted to be negative ($-$), 
\begin{equation}
\frac{|\{x \in S^- : v(x)< \theta\}|}   {|S^-|}~.
\end{equation}

For the prediction of lethality, $S$ is the set of all single knock-outs in all
network realizations considered, $S^+$ are the lethal knock-outs, $S^-$ the viable
ones. Analogously in the context of double knock-outs, $S^+$ are the synthetically
lethal cases. The function $v$ is the predictor used, such as out-degree, in-degree etc.
ROC plots show sensitivity as a function of specificity subtracted from 1.

\paragraph{Acknowledgments}
This work has been funded by VolkswagenStiftung through the initiative on
Complex Networks as Phenomena across Disciplines.
\bibliographystyle{mde}
\bibliography{synlet}
\end{document}